# Categorization of Next Generation Nanomaterials: International Cooperation on the Categorization of Nanomaterials in the Regulatory Chemical Context

**Maria J. Doa**

United States Environmental Protection Agency, Office of Research and Development, Washington DC, USA

Email: doa.maria@epa.gov

ORCID iD: 0000-0002-5282-8016

**Abstract**

Categorization approaches have been effectively applied to chemicals, and many have tried to apply variations of these approaches to nanomaterials. Given the added complexities of nanomaterials, this has been challenging. International cooperation on categorization approaches has been made, primarily through the Organization for Economic Cooperation and Development. Progress has been limited given the complexities of nanomaterials, especially given the need to consider not only the intrinsic properties of the materials but also properties dependent on the system into which it is introduced. Consideration must also be given to the different purposes and contexts, in particular regulatory contexts, to which the categorization schemes would be applied. More progress can be anticipated by focusing on areas of overlap among countries such as physicochemical properties.

**Keywords**: categorization of nanomaterials, chemical characterization, engineered nanomaterials, international cooperation, nanomaterials

**Introduction**

Categorization is useful because it organizes information about a group of substances that can be extrapolated to other substances that fall within certain specific parameters, e.g., molecular weight or presence of certain functional groups. The usefulness of categorization and grouping approaches has been recognized [1]; and these approaches are used in regulatory contexts for chemicals because the use of categorization approaches provides for targeted and streamlined regulatory reviews, targeted health and safety testing, and better regulatory decision making. Another benefit is that these approaches communicate potential concerns with a type of chemical or material and identifies testing that would be required to inform that concern.

Categorization of nanomaterials would bring many of the benefits in the regulatory context associated with the use of categories for chemicals. However, applying this construct to nanomaterials is more challenging because of the added complexities of nanomaterials as compared to chemicals. Given that many nanomaterials have been specifically engineered into nanoscale structures that have particular properties, the hazards associated with nanomaterials may go beyond the chemical make-up of the nanomaterial and can be associated with the material's other attributes. Some of the properties are intrinsic to the nanomaterial, and others are dependent upon the system into which the nanomaterial is placed or into which it is released. This makes nanomaterials difficult to categorize. Despite these additional complexities, for the purposes of regulatory assessment, chemical categories can provide a starting point for the consideration of nanomaterial categorization, particularly for hazard analogue-based regulatory frameworks.

**Chemical Categorization**

Chemical categories are developed when there are adequate data of sufficient quality about a group of similar chemicals that would support general applicability to other potential members of the category based on physicochemical properties, and human health, environmental toxicity and/or environmental fate. A category is defined by specifying the molecular structure a new chemical must have to be included in it. Properties such as molecular weight, equivalent weight, water solubility, common mode of mechanism, or outcome pathway are often used to further determine inclusion in (or exclusion from) a category. Within the category, there should be incremental and constant change of parameters (such as hazard endpoint and physicochemical properties) across the category. Finally, specific standard hazard and fate tests are often identified for the category to address concerns for the category (e.g. in reference [2]). Categories include chemicals for which sufficient history has been accumulated so that hazard concerns and testing recommendations vary little from chemical to chemical within the category.

Several jurisdictions use chemical categories. The United States Environmental Protection Agency (EPA) uses hazard-based categories as an aid in evaluating chemicals [2], typically in evaluating chemicals for which there are limited data. When a chemical is identified as a member of a category (e.g., based on structural similarities), it is evaluated in the context of the potential health or environmental concerns associated with that category. Categorization can help streamline the regulatory decision-making process, and when used across jurisdictions (e.g., Canada and the United States) can simplify the regulatory landscape. Categories are a useful tool to anticipate the type of regulatory controls that could be applied to that particular category of chemicals, e.g., limitations on releases to surface waters, or the minimization of worker exposure. Thus, when developing a chemical, a company can use the information on the category as a screen in the design process.

**Categorization of Nanomaterials**

While the categorization of chemicals, particularly for purposes of hazard assessment and characterization of persistence and bioaccumulation, is used in many jurisdictions, adoption of similar categorization for nanomaterials has proved more difficult. Additional challenges in using such a framework for nanomaterials are due to intrinsic and extrinsic differences in physical and chemical properties and differences among nanoscale forms (*nanoforms*) of a chemical and between nano and non-nano forms of a chemical. Further, nanomaterials often do not exist as distinct species; rather the populations of the materials can consist of distinct species and agglomerates and aggregates. Thus, in looking at how to group nanomaterials and in distinguishing among them, in addition to chemical composition, surface chemistry and shape and morphology, the following physicochemical properties are important in a regulatory context (e.g. [3]), adding complexity to the exercise: particle size and particle size distribution, surface charge, specific surface area, and dispersion stability. Nanomaterials are generally materials engineered to have specific properties, and thus are different from bulk chemicals, whose properties are functional group-based (e.g., ketone) or metal-based (e.g., hexavalent chromium). In a regulatory context, regulators, particularly in North America, typically distinguish substances under their respective laws based on a molecular identity (material) as opposed to only a "*properties*" basis. Thus, in developing a categorization scheme, chemical identity is likely to be an important component of such a scheme. However, to take into account the nature of nanomaterials, a categorization scheme based on chemical categorization will also be a hybrid that is dependent to a much larger degree than chemical categories on the common physicochemical properties. This introduces a weakness of such categorization schemes given limitations in the consistency and quality of physicochemical measurements for nanomaterials.

A range of approaches to categorization for nanomaterials have been proposed, including schemes based primarily on physicochemical properties, type of adverse effect, or chemical constituents. These different approaches are due in part to the complexities of the nanomaterials themselves; to the different purposes of the categorization, e.g., risk screening hazard grouping; and in part to whether its use is in a regulatory context. A few examples of this range follow.

- Kuempel et al [4] discussed hazard and risk-based categorical approaches to occupational exposure control using respirable poorly soluble particulates as a case study. They investigated the development of occupational exposure limits for categories defined by mode of action (MOA) and a benchmark particle. A constraint on the breadth of the category is the potency of the MOA.

- Oomen et al [5] discussed grouping approaches for the risk assessment of nanomaterials, based on a core set of physicochemical properties that focused on testing strategies and the "relationships between the variable" physicochemical properties and hazard, toxico-kinetics, fate, exposure. This approach examined "what are the physical and chemical identities of the nanomaterials," "where they go" based on solubility, hydrophobicity, dispersability, and dustiness; and "what they do" based on physical hazards, biological (re)activity and photoreactivity.

- Tervonen et al [6] proposed a risk-based categorization scheme focused on environmental risk. The grouping for individual nanomaterials considered physicochemical properties, the environmental fate parameters of bioavailability and bioaccumulation and toxicity weighing. This approach allows for initial risk screening of individual types of nanomaterials considered for use in consumer products.

- In the United States, a chemical-based scheme is being considered because it is more consistent with the chemical regulatory framework used to evaluate nanomaterials under the Toxic Substances Control Act (TSCA) [7]. Indeed, EPA has assessed about 200 nanomaterials under TSCA using to some extent grouping based on chemical constituent. The majority are carbon-based nanomaterials, quantum dots, and coated metal or metal oxide structures. EPA is developing a framework for categorization of nanomaterials based on chemical constituent and physicochemical properties [8]. This approach is used in tandem with an existing chemical substance category. For human health, in the absence of data on the specific nanomaterial, EPA groups nanomaterials broadly under the category of respirable poorly soluble particulates (this category is limited to effects on the lung as a result of inhaling the particles; effects range from inflammation to fibrosis to, potentially, cancer) [2] with a separate consideration for any toxicity associated with the constituent chemicals.

Even among schemes based on chemical constituents, there are variations in the degree of categorization. For example, some schemes group all carbon-based nanomaterials together, e.g., carbon nanotubes, graphene, carbon black. Others treat these separately or even distinguish further, such as between multiwall carbon nanotubes and single and double-wall carbon nanotubes. For example, Stone et al [9] considered both shape and size as a basis for one approach and broad groups based on chemical constituent as a second option for categorization for environmental studies: carbon (carbon black, nanotubes and fullerenes), mineral-based (metals, metal oxides), organic (polymers, dendrimers, surfactant coatings) and composites/hybrids (multicomponent nanomaterial, e.g., quantum dots, or doped metal/metal oxides). Physicochemical factors were considered in a second tier.

Whether the categorization frameworks are hazard, exposure, risk, or novel-property based, they generally include physicochemical properties as an integral component. Different categorization schemes use similar but different sets of physicochemical properties. The physical and the chemical properties that are commonly considered as fundamental to a categorization are shape, particle size, specific surface area, surface charge, dispersibility.

These examples give a flavor for the range of categorization approaches that have been considered. Some of these have been considered outside of a regulatory context while others are being developed within different regulatory contexts (e.g., [5, 7]). The regulatory context is important because it is a significant factor in determining the appropriate categorization scheme. Thus, while a chemical component approach is consistent with the regulatory construct in some jurisdictions, such as the United States, it is not necessarily fit for purpose in the regulatory context in other jurisdictions. This raises the question of how much international cooperation is ultimately feasible given the differences in regulatory constraints in different jurisdictions. Some jurisdictions are limited to a primarily chemical-based approach while other jurisdictions have greater flexibility to consider other factors such as dispersibility and biological reactivity as equivalent to chemical makeup in categorizing nanomaterials (for example, see Oonen et al [5]).

**International Cooperation Through the Organization for Economic Cooperation and Development**
The importance of international cooperation in the categorization of nanomaterials has been recognized by the member countries of the Organization for Economic Cooperation and Development (OECD). The OECD Working Party for Manufactured Nanomaterials (WPMN) has supported efforts to ensure that approaches for hazard, exposure and risk assessment for manufactured nanomaterials are of high quality, science-based and internationally harmonized. The OECD and its member countries have determined that the approaches for the testing and assessment of chemicals are in general appropriate for assessing the safety of nanomaterials but may have to be adapted to nanomaterials [10].

To assess the problems associated with the complexities of nanomaterials, the OECD WPMN held two expert meetings on the categorization of nanomaterials (OECD 2015, 2016). The first expert meeting included scientists, regulators, industry representatives, and non-governmental organization stakeholders. The scope and overarching goals of the expert meetings were broad: to develop and define categorization of nanomaterials. Grouping schemes were proposed for testing, individual hazard endpoints, environmental fate, read across/structure activity relationships (SARs), and exposure. The proposed approaches represented a wide range of categorization types, solely physicochemical property based, exposure based, hazard based, risk-based, novel properties-based [11].

While the meeting did recognize the importance of categorization in many aspects of health and safety, it did not propose a general framework for categorization. There was recognition of the importance of developing a scheme that is fit-for-purpose and supported adapting *existing* frameworks for conventional substances to fit categorization frameworks for nanomaterials.

A common thread throughout the meeting was the important role that physicochemical properties play in all categorization schemes for nanomaterials. Participants noted the critical need for reliable physicochemical data in characterizing properties inherent to the parent nanomaterial and properties dependent on the different systems in which the nanomaterial may exist, e.g., air, soil [11].

The second meeting focused on identifying common denominators among different approaches and frameworks for grouping nanomaterials that could be used in regulatory context. It addressed different regulatory schemes and their influence on the parameters of categorization schemes. There were no recommendations for specific categorization schemes. It was concluded, however, that the general scheme for building and justifying groups and the read-across seems acceptable for nanomaterials, but there is a need to develop guidance that addresses the specificities of nanomaterials [12].

A common theme throughout these discussions is the importance of physicochemical properties in a categorization scheme for nanomaterials, which is probably also the most promising area for synergies among different regulatory systems. A related and somewhat confounding issue is the challenge in developing reliable data using methods that are appropriate for a specific physicochemical property and a specific type of nanomaterial. Once a categorization scheme is chosen, a fundamental aspect of a category is the determination of the variability of the physicochemical properties among the specific constituents of the category and the trends in those properties with varying attributes, such as size and shape and surface functionalization. As with many aspects of nanotechnology, the quality of the physicochemical data is the keystone to credible categorization. This is an area for which there is the opportunity for greater cooperation. Physicochemical properties are key to all categorization schemes.

The OECD has recognized this and has held two additional expert meetings (one in collaboration with ISO TC 229) on physicochemical properties [13, 14] and conducted an evaluation of the data developed under the WPMN Sponsorship Testing Programme [15]. This has supported work within the OECD WPMN on developing guidance on the appropriate characterization of nanomaterials and identifying which test guidelines should be developed or adapted for specific physicochemical endpoints. Based on these meetings and the WPMN Sponsorship Testing Programme data, the Netherlands and the United States, in conjunction with other OECD member countries, are leading the development of a categorization-based decision tree for physicochemical property testing. This framework would include decision trees for each physicochemical endpoint. Each decision tree would identify, based on the specific type of nanomaterial and the type of assessment, the appropriate method(s) to be used for a physicochemical endpoint. The methods that are not considered appropriate for specific nanomaterials for a particular purpose (e.g., for use only in screening or in a more robust risk assessment) would also be identified. The development of the categorization-based decision tree would be used to support the other types of categorization being used for nanomaterials. This categorization of physicochemical properties could then be used to support different toxicity, environmental fate, risk categorization schemes that jurisdictions development to meet their specific needs.

**Conclusion**
Categorization is an approach that increases efficiencies and is useful in reviewing both individual and groups of substances. Given the large number of nanomaterials, variations among nanomaterials, and the limited number of nanomaterials with robust data sets, categorization can be used to improve decision-making, whether in the regulatory context, designing products, or for establishing research priorities. The type of categorization scheme - physicochemical-based or chemical-based - in conjunction with other factors, e.g., bioreactivity, dispersibility will depend on the context of its use, particularly the regulatory context and the need the categorization scheme will address.

There are opportunities for broad cooperation on the categorization of nanomaterials in areas such as physicochemical properties. International fora such as the OECD will continue to be important in highlighting these opportunities. There will also continue to be international cooperation on the toxicity characterization and risk assessment of nanomaterials. However, given that some jurisdictions may prefer or are constrained to applying the constructs developed for chemicals to the categorization of nanomaterials, this may limit the ability to develop internationally consistent nanomaterial categorization schemes for toxicity and risk assessment in the same manner as has been done for chemicals.

**Acknowledgements**
This article is one of a collection of articles about the categorization of nanomaterials, generated by research and workshop discussions under the FutureNanoNeeds project funded by the European Union Seventh Framework Programme (Grant Agreement No 604602). For an overview and references to other articles in this collection, see *The Nature of Complexity in the Biology of the Engineered Nanoscale Using Categorization as a Tool for Intelligent Development* by Kenneth A. Dawson.

Author declares there is no conflict of interest.